\documentclass[aps,prb,twocolumn,groupedaddress,epsfig]{revtex4}

\usepackage{epsfig}
\usepackage{graphics}

\begin{document}

%\preprint{}

\title{Enhancement of spontaneous emission in a quantum well by resonant
 surface plasmon coupling}


\author{Arup \surname{Neogi}, Chang-Won \surname{Lee}, and Henry O. \surname{Everitt}}
\email[Email address]{: everitt@aro.arl.army.mil}
%\thanks{}
\affiliation{Department of Physics, Duke University, Durham, NC 27708, USA}
\author{Takamasa \surname{Kuroda} and Atsushi \surname{Tackeuchi}}
\affiliation{Department of Applied Physics, Waseda University,
Okubo 3-4-1, Shinjuku, Tokyo 169-8555, Japan}
\author{Eli \surname{Yablonovitch}}
\affiliation{Department of Electrical Engineering, University of
California, Los Angeles, CA  90095, USA}

\date{\today}

\begin{abstract}
Using time-resolved photoluminescence measurements, the
recombination rate in an In$_{0.18}$Ga$_{0.82}$N/GaN quantum well
(QW) is shown to be greatly enhanced when spontaneous emission is
resonantly coupled to a silver surface plasmon. The rate of
enhanced spontaneous emission into the surface plasmon was as much
as 92 times faster than normal QW spontaneous emission. A
calculation, based on Fermi's golden rule, reveals the enhancement
is very sensitive to silver thickness and indicates even greater
enhancements are possible for QWs placed closer to the surface
metal coating.
\end{abstract}

%\keywords{}

\maketitle

The spontaneous emission (SE) decay constant $\tau$ for radiating
dipoles at $\vec{r}_{e}$ is given by Fermi's golden rule
\begin{equation}
\frac{1}{\tau}=\frac{2\pi}{\hbar}|\langle
f|\vec{d}\cdot\vec{E}(\vec{r}_{e})|i\rangle|^{2}\rho(\hbar\omega),
\end{equation}
where $\rho(\hbar\omega)$ is the photon density of states (DOS)
and $\langle f|\vec{d}\cdot\vec{E}(\vec{r}_{e})|i\rangle$ is the
dipole emission matrix element. As pointed out by Purcell, SE may
be enhanced by altering the photon DOS\cite{PurcellPR69}. For
example, the ratio of enhanced to free space emission (the Purcell
factor $F$) has been measured as large as 5 in an atomic system by
placing the radiating atoms in a high $Q$, low volume
cavity\cite{GoldsteinSpontaneous,DutraPRA53}. A Purcell factor of
up to 6 has been observed from quantum well (QW) and quantum dot
emitters in vertical cavity surface emitting laser structures,
while an enhancement of 15 has been observed from quantum dots in
a microdisk cavity\cite{WeisbuchJL85,GerardPRL81}. Photonic
crystals and distributed Bragg gratings have also been used to
enhance the SE rate by as much as a factor of
4.5\cite{VuckovicJQE36,BoroditskyJLT17,BabaJLT17}. Such enhanced
SE rates, achieved by increasing the photonic DOS in a small
cavity, permit lower threshold, higher modulation frequency lasers
as well as more efficient light emitting diodes.

The SE rate can also be modified when semiconductor or dye
emitters are coupled to a surface plasmon (SP) of a metallic film
\cite{HeckerAPL75,BabaJJAP35,GontijoPRB60,BrillianteJMS79}. A
single QW can experience strong quantum electrodynamic coupling to
a SP mode if placed within the SP fringing field penetration
depth. An electron-hole pair in the QW recombines and emits a
photon into a SP mode instead of into free space. The degree of SE
rate modification for a given wavelength depends on the SP DOS at
that wavelength. The strongest enhancement occurs near the
asymptotic limit of the SP dispersion branch, the SP "resonance"
energy $E_{sp}$, where the SP DOS is very high. Non-resonant,
SP-mediated SE enhancements as large as 6 have been observed from
GaAs QWs near thin Ag films\cite{HeckerAPL75}. Even greater
enhancements are possible for wide bandgap semiconductors whose
emission wavelength is coincident with $E_{sp}$. In this report,
time-resolved photoluminescence (TRPL) measurements of a partially
silver-coated InGaN/GaN QW directly demonstrate the SP-mediated
resonant enhancement of the SE rate for the first time in a
semiconductor QW.

An InGaN/GaN QW was used in these experiments, grown by
metal-organic chemical vapor deposition (MOCVD) on sapphire
substrate\cite{KellerJCG195}. Over a 1.5 $\mu$m Si-doped GaN
buffer layer was grown a 28 nm In$_{0.04}$Ga$_{0.96}$N reference
layer, a 6 nm GaN layer, and the 3 nm In$_{0.18}$Ga$_{0.82}$N QW
as shown in Fig. 1. Above the QW was a 12 nm Si-doped GaN cap
layer, placing the QW within the fringing field depth of the SP. A
layer of silver, $\sim$ 8 nm thick, was deposited by electron beam
evaporation on one-half of the sample surface. The other half was
left bare to facilitate direct comparison of the silvered and
unsilvered results.

\begin{figure}
\resizebox{6.2cm}{!}{\hbox{\includegraphics{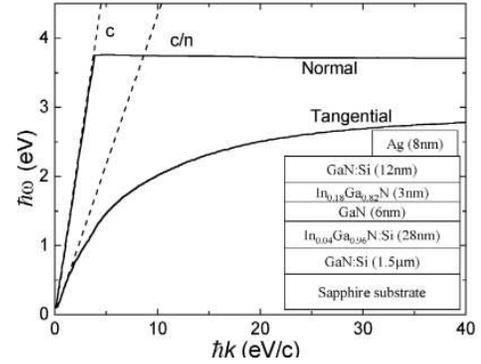}}}
\caption{The calculated surface plasmon dispersion relation for
tangential and normal modes.
 Inset: The structure of the sample studied.}
\label{Fig1}
\end{figure}
The bulk plasmon energy of silver is 3.76 eV, but the SP energy of
Ag is lowered by the GaN dielectric
constant\cite{EhrenreichPR128}. The SP dispersion relation is
derived, using Maxwell's equations, from the known dielectric
properties of Ag and GaN\cite{BrunnerJAP82}. Considering a silver
film of thickness $t$ and permittivity $\varepsilon_{2}$,
sandwiched between GaN and air with permittivities
$\varepsilon_{1}$ and $\varepsilon_{3}$ respectively, the boundary
condition gives the SP dispersion relation
\begin{equation}
    (\frac{\gamma_{1}}{\varepsilon_{1}}+\frac{\gamma_{2}}{\varepsilon_{2}})
    (\frac{\gamma_{3}}{\varepsilon_{3}}+\frac{\gamma_{2}}{\varepsilon_{2}})-
    (\frac{\gamma_{1}}{\varepsilon_{1}}-\frac{\gamma_{2}}{\varepsilon_{2}})
    (\frac{\gamma_{3}}{\varepsilon_{3}}-\frac{\gamma_{2}}{\varepsilon_{2}})e^{-2\gamma_{2}t}=0,
\end{equation}
where $\gamma_{i}=k^{2}-\varepsilon_{i}\omega^{2}/c^{2}$ , $i=
1,2,3$, and $k = 2\pi/\lambda$ is the wavevector. The SP
dispersion contains tangential and normal mode branches (Fig. 1),
indicating the dominant direction of current flow in the silver
film. For silver films with $t\geq$ 8 nm, the tangential SP branch
asymptotically approaches $E_{sp}=$ 2.85 eV ($\lambda_{sp}=$ 436
nm), the SP "resonance" energy. Because the photon DOS is
proportional to $dk/d\omega$, the SP DOS and SE enhancement will
be greatest at the SP resonance.

The resonant enhancement is measured by comparing the luminescence
decay rate from the photoexcited QW on the silvered and unsilvered
sides. Room temperature TRPL measurements were performed using a
100 MHz Kerr-lens mode-locked, frequency-doubled Ti:Sapphire(Ti:S)
laser with average incident pump power of 10 mW ($\sim$ 13
$\mu$J/cm$^{2}$). The pump excitation energy (3.14 eV or 395 nm)
was chosen to be below the bandgap of the InGaN reference layer
and GaN layers so electron-hole pairs were generated only in the
QW. The luminescence signal was dispersed in a grating
spectrometer (600 gr/mm) and measured simultaneously across three
wavelength bands (2.55-2.68, 2.71-2.87, and 2.79-2.96 eV) using a
Hamamatsu streak camera with a resolution of 15 ps. The three
adjacent 25 nm ($\sim$ 150 meV) wide wavelength bands spanned the
entire continuous wave photoluminescence (cw PL) emitted from the
QW. Features narrower than 25 nm were resolved by sequentially
comparing adjacent 5 or 10 nm data windows offset from each other
in 1 - 4 nm steps. Of course, all TRPL traces represent the sum
effect of components with wavelength dependent behavior. However,
narrowing the bandwidth of the windows further did not
significantly improve the ability to measure wavelength dependence
because of the reduced signal-to-noise, especially on the silvered
side.

\begin{figure}
\resizebox{6cm}{!}{\hbox{\includegraphics{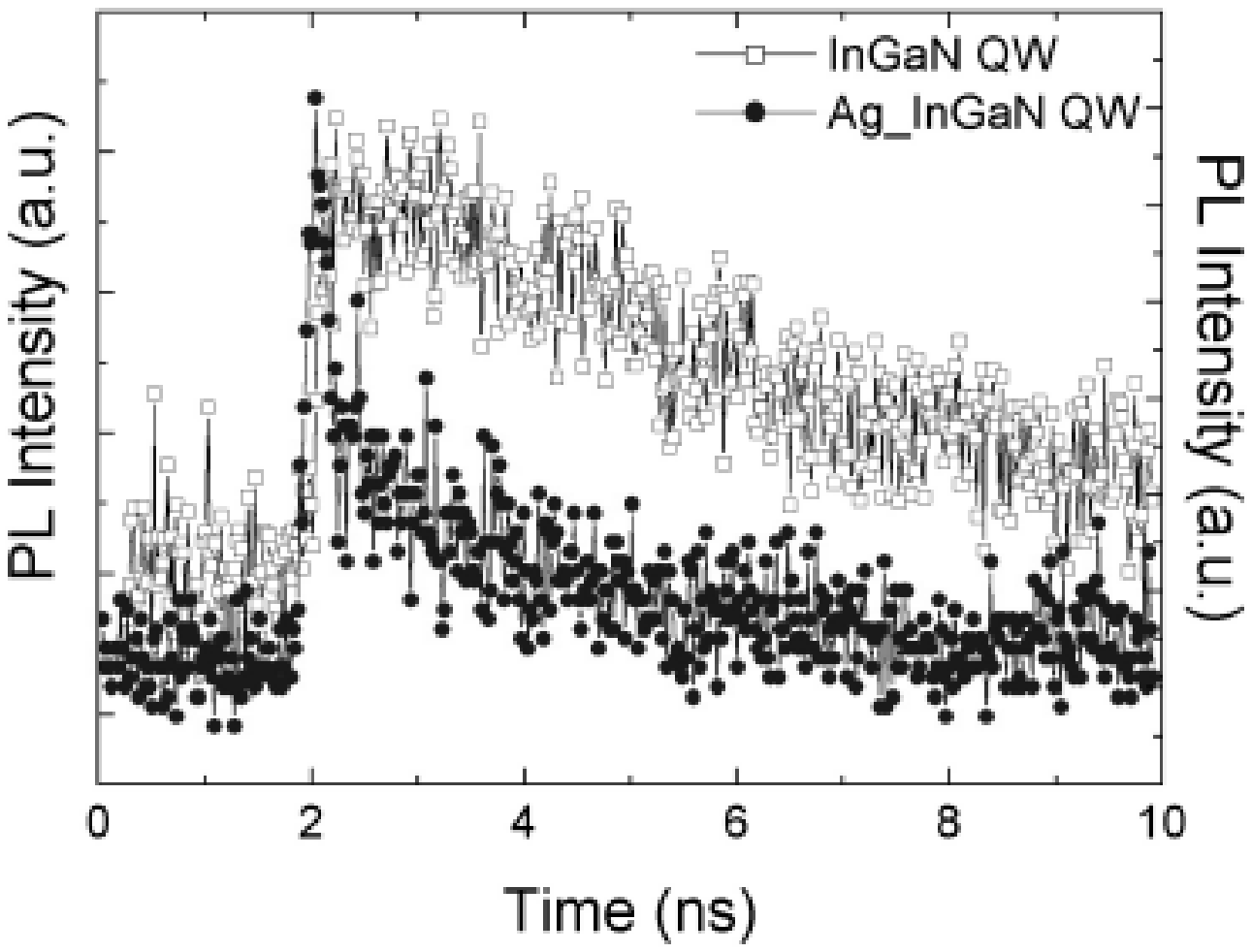}}}
\resizebox{6cm}{!}{\hbox{\includegraphics{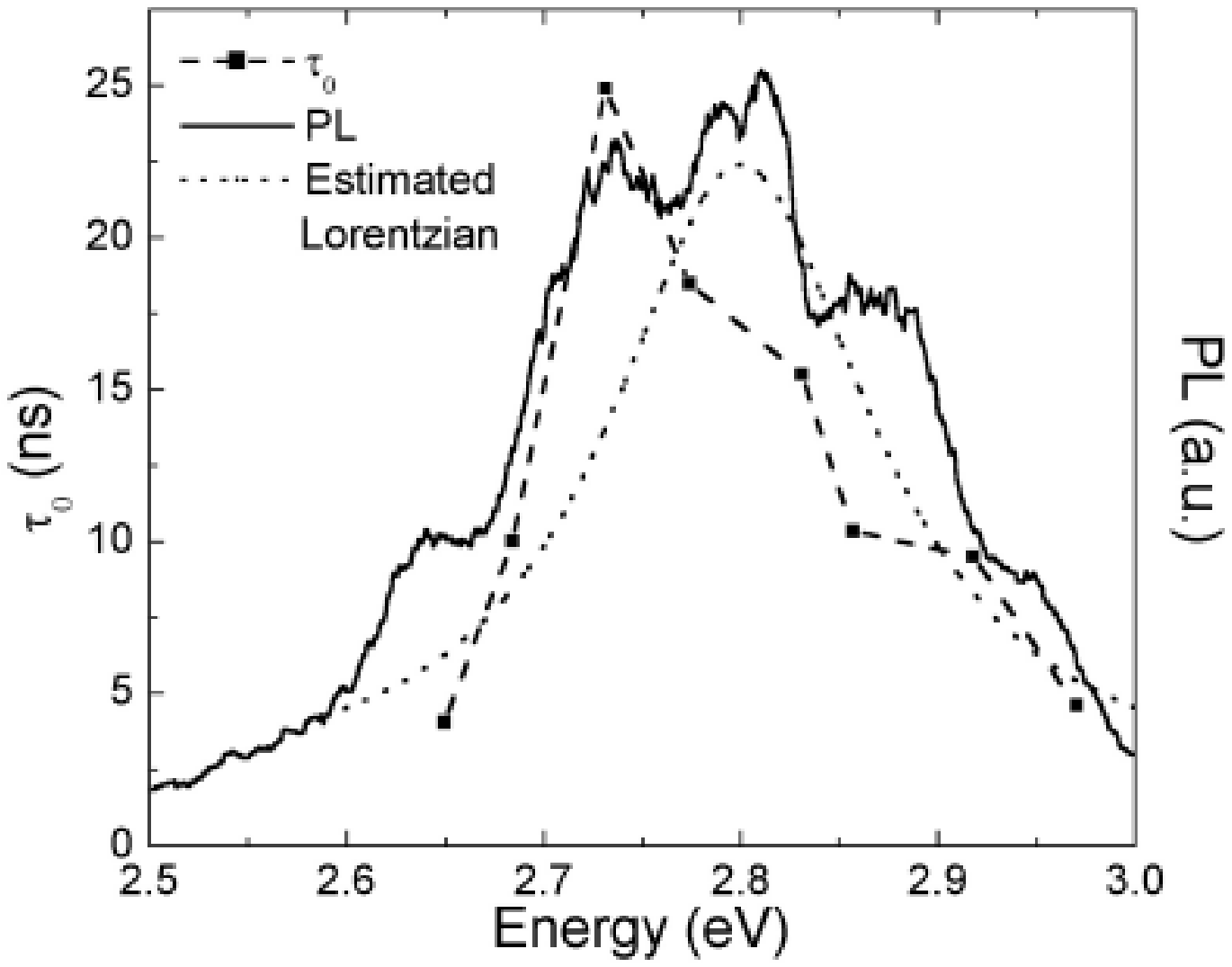}}} \caption{a)
TRPL decay of the unsilvered and silvered InGaN QW for a 25
nm-wide wavelength detection window (2.79-2.96 eV), with pump
energy of 3.14 eV. b) Comparison of the cw PL and the
TRPL-measured recombination rate
 constant ($\tau_{0}$) of the unsilvered InGaN QW.
 The dashed curve is the estimation given by (4)with $\hbar \omega_{c} =$
  2.8 eV and $\hbar \Delta \omega =$ 0.16 eV.}
\label{Fig2}
\end{figure}
An example of a TRPL trace is presented in Fig. 2a, comparing the
temporal decay of the unsilvered and silvered QW PL for a 25
nm-wide wavelength band. The luminescence decay constants from the
silvered and unsilvered sides were independently derived from
exponential fits to the data (5 or 10 nm windows) and then
compared under identical pump and detector parameters. The
uncertainty in fitted rate constants for a 10 nm window is 2 - 5
\% at wavelengths near the peak cw PL emission (2.75 eV) and rises
to as much as 10 - 16 \% at longer and shorter wavelength extremes
where the emission is weaker.

On the unsilvered side, the QW exhibited a long single exponential
decay whose decay constant $\tau_{0}$ was slowest ($\tau_{0}=$ 28
ns) at wavelengths near the peak PL emission (2.76 eV) (Fig. 2b).
This very long decay constant, and the correspondingly high PL
intensity, indicate a high internal quantum efficiency and
insignificant non-radiative (nr) processes compared to radiative
(r) recombination ($1/\tau_{0} = 1/\tau_{nr}+ 1/\tau_{r}\simeq
1/\tau_{r}$) \cite{MonemarSPIE3624,ChichibuCh5}. Away from the
peak PL emission wavelength, the recombination rate accelerates
($\tau_{0}=$ 4 - 5 ns) at the longest and shortest wavelengths
measured.  The wavelength dependence of QW emission is well
understood from earlier TRPL studies of InGaN QWs
\cite{ChichibuCh5} but differs markedly from the SE rate of a
dipole radiator (dipole moment $d$) in a dielectric medium (index
of refraction $n$)
\begin{equation}
    \frac{1}{\tau_{r}(\omega)}=\frac{4nd^{2}\omega^{3}}{3\hbar c^{3}},
\end{equation}
especially at low frequency where shallow level traps, impurities,
and quantum dot-like structures in the QW contribute to
recombination.  A phenomenological estimate is a Lorentzian
\begin{equation}
\frac{1}{\tau_{0}(\omega)} \simeq \frac{1}{\tau_{r}(\omega)}=
\frac{1}{\tau_{r_{0}}}\frac{\Delta\omega^{2}}{(\omega-\omega_{c})^{2}+\Delta\omega^{2}},
\end{equation}
whose peak at ($\omega_{c}$) and linewidth $\Delta\omega$ also
roughly coincide with that of the measured cw PL (Fig. 2b).

\begin{figure}
\resizebox{6.5cm}{!}{\hbox{\includegraphics{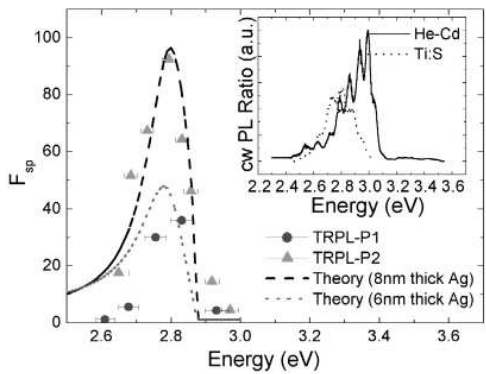}}}
\caption{The Purcell enhancement factor, $F_{sp}$, measured using
TRPL windows 10nm (position $P_{1}$) and 5nm (position $P_{2}$)
wide. Overlaid is the prediction of the enhancement. Inset:
Comparison of the time-integrated PL emission ratios for
excitation by HeCd and Ti:S lasers.} \label{Fig2}
\end{figure}

By contrast, the weaker PL intensity through the silver-coated
surface exhibits a bi-exponential decay. The faster decay
component, with decay constant $\tau_{1}$, corresponds to enhanced
recombination mediated by the SP mode
($1/\tau_{1}=1/\tau_{nr}+1/\tau_{r}+1/\tau_{sp}\simeq
1/\tau_{sp}$). As expected, this faster component is strongest
near $E_{sp}$ (2.85 eV), but it is evident at energies 200meV
lower than $E_{sp}$  and decays with an almost
wavelength-independent constant of $\tau_{1}\simeq \tau_{sp}\sim$
235 ps.  Evidently, the SP resonance is fairly broad. The slower
TRPL relaxation component has a decay constant ($\tau_{2}$) which
decreases with decreasing wavelength from 5 ns at 2.94 eV to 9 ns
at 2.61 eV.  This component is the remnant radiative relaxation
component of the QW ($\tau_{r}$).  The frequency dependence is
approximately $\omega^{-3}$ as in (3), and the role of other
nonradiative processes is not evident.

To summarize the wavelength dependence of the SP enhancement, Fig.
3 plots the Purcell factor ($F_{sp}=1+\tau_{0}/\tau_{1}$) derived
from the measured TRPL decay constants. The data demonstrate a
sudden rise at higher energies, a peak enhancement at 2.8 eV, and
weaker enhancement at lower energies. The maximum values of
$F_{sp}$ were 36 (at 2.83 eV) and 92 (at 2.79 eV) at sample
positions $P_{1}$ and $P_{2}$, respectively. By comparing the
relative cw PL intensities of the InGaN reference layers, it is
evident that the silver film is thicker at $P_{2}$ than at
$P_{1}$. The differing $F_{sp}$ values at $P_{1}$ and $P_{2}$
consequently suggests a sensitivity of $F_{sp}$ to Ag thickness.

Fermi's golden rule (1) provides insight into the frequency
dependence of the Purcell factor and reveals the sensitivity of
$F_{sp}$ to the silver thickness $t$ and Ag-QW separation
$a$\cite{GontijoPRB60}. First, the electric field of the SP mode
at the QW must be calculated and used to derive the dipole matrix
element. The SP electric field varies only in the $z$ direction,
so the normalization of $E(z)$ to a half quantum of zero-point
fluctuation in the dispersive medium becomes
\begin{equation}
\alpha^{2}=\frac{S}{A}=\frac{\hbar\omega/2}{\frac{A}{8\pi}\int_{-\infty}^{\infty}dz
    \frac{\partial(\omega\varepsilon(\omega,z))}{\partial\omega}|E(z)|^{2}},
\end{equation}
where $E(z)$ is the unnormalized electric field at a distance $z$
from the Ag-GaN interface, $|\alpha E(a)|^{2}$ is the normalized
electric field at QW depth $a$, $A$ is the quantization area, and
$\varepsilon(\omega,z$) is the dielectric function of the GaN, Ag,
or air. The enhanced recombination rate ($1/\tau_{sp}$) can then
be estimated in the QW under the influence of the local electric
field from the tangential SP mode
\begin{eqnarray}
    \frac{1}{\tau_{sp}(\omega)}=
    \frac{2\pi}{\hbar}(\frac{1}{3}d^{2}|\alpha E(a)|^{2})2\pi k
    \frac{A}{(2\pi)^{2}}\frac{dk}{d(\hbar\omega)}\nonumber
\end{eqnarray}
\begin{eqnarray}
=\frac{S}{3\hbar^{2}}|dE(a)|^{2}k\frac{dk}{d\omega},
\end{eqnarray}
where the factor $1/3$ comes from spatial averaging of
polarization. Inserting measured values for $a$ (12 nm) and $t$ (8
nm) while using the $\tau_{1}$ data to fit for $d$ (24 nm), the
calculation confirms that $\tau_{sp}$ is relatively independent of
frequency from 2.6 eV to $E_{sp}$.

The frequency dependence of $F_{sp}$ derives from (6) and the
frequency dependence of the unsilvered QW recombination rate
$\tau_{0}$ (Fig. 2b). For the parameters in this experiment, the
frequency dependence of $F_{sp}$ below $E_{sp}$ derives primarily
from $\tau_{0}$. Unfortunately, a predictive theory of the
$\tau_{0}$ is not available, so an accurate estimate of $F_{sp}$
is not possible. However, if (4) and (6) are used as
approximations for $\tau_{0}$ and $\tau_{1}$, respectively, then
$F_{sp}(\omega)\simeq 1+\frac{\tau_{0}}{\tau_{1}}$ may be plotted
(Fig. 3) using the parameters for this sample. The predicted
frequency peak and width of $F_{sp}$ agree well with the measured
peaks and widths of the TRPL data.  The peak value of $F_{sp}$ is
a sensitive function of Ag thickness through $\tau_{sp}$.  A
reduction of $t$ from 8 to 6 nm halves the peak value of $F_{sp}$,
suggesting that the differing values of $F_{sp}$ observed at
$P_{1}$ and $P_{2}$ arise from small variations of the Ag
thickness.

Recently, $F_{sp}$ was measured in a similar sample by means of cw
PL using a HeCd excitation source ($E_{ex}=$ 3.82 eV,
$\lambda_{ex}=$ 325 nm) \cite{GontijoPRB60}. A ratio of the PL
emission for the unsilvered to silvered sides was measured as a
function of wavelength, and a peak enhancement factor of 55 was
estimated from the data. Those measurements were repeated here,
and although the measured enhancement factors were similar to and
consistent with the TRPL data, the energy of maximum enhancement
in both cw PL measurements was blue shifted from $E_{sp}$. When
time-integrated PL is measured using the Ti:S laser instead, this
blue shift disappears (inset, Fig. 3). Note that the HeCd laser
excites all layers of the sample while the Ti:S laser only excites
the QW. Therefore, any cw PL ratio measured using a HeCd laser
must be understood to represent $F_{sp}$ convolved with other
excitation-dependent effects (particularly in the cap layer),
while the TRPL data measures only $F_{sp}$.  Furthermore, the role
of nonradiative recombination was ignored in that work, leading to
an overly simiplistic estimate of $F_{sp}$.

\begin{figure}
\resizebox{6cm}{!}{\hbox{\includegraphics{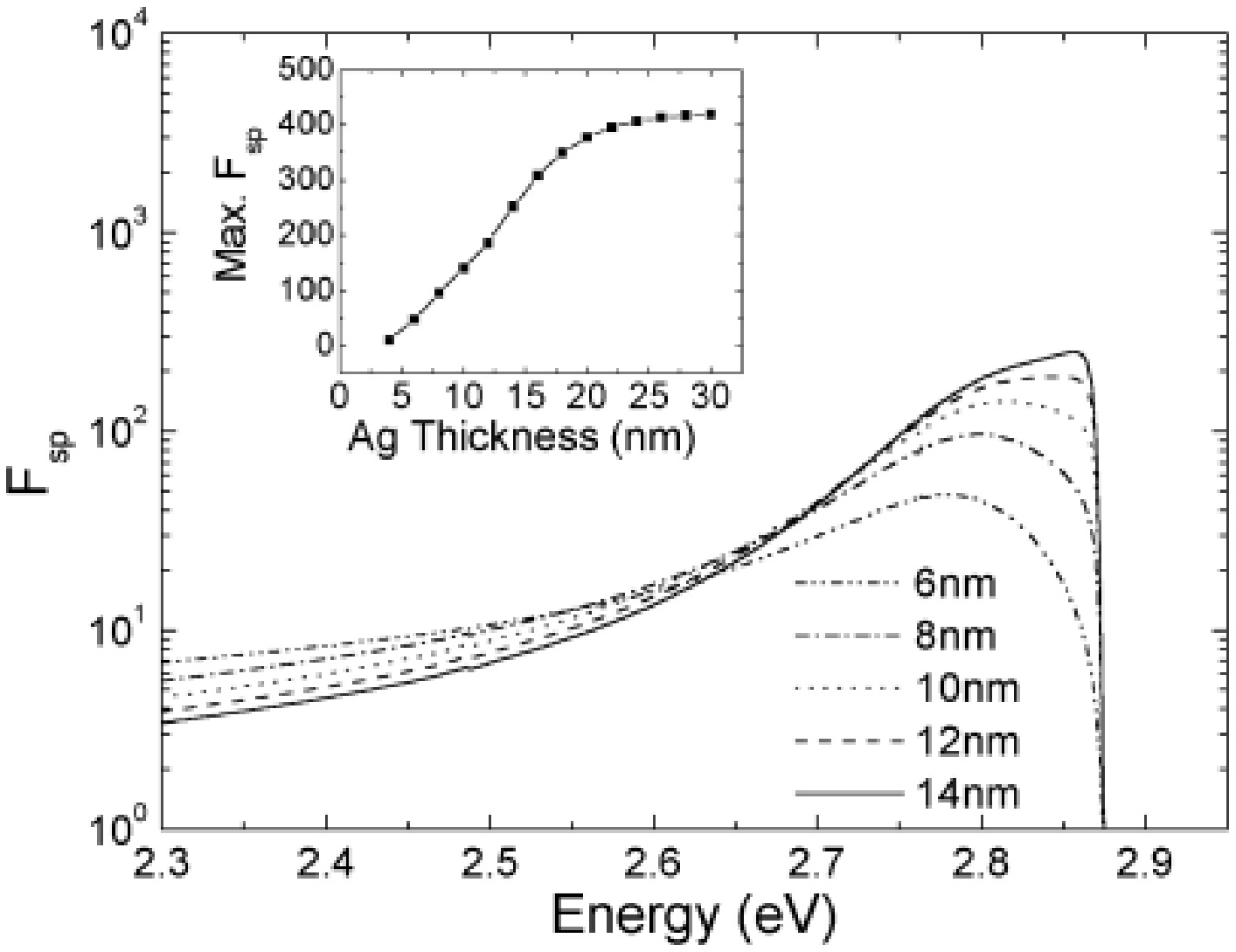}}}
\resizebox{6cm}{!}{\hbox{\includegraphics{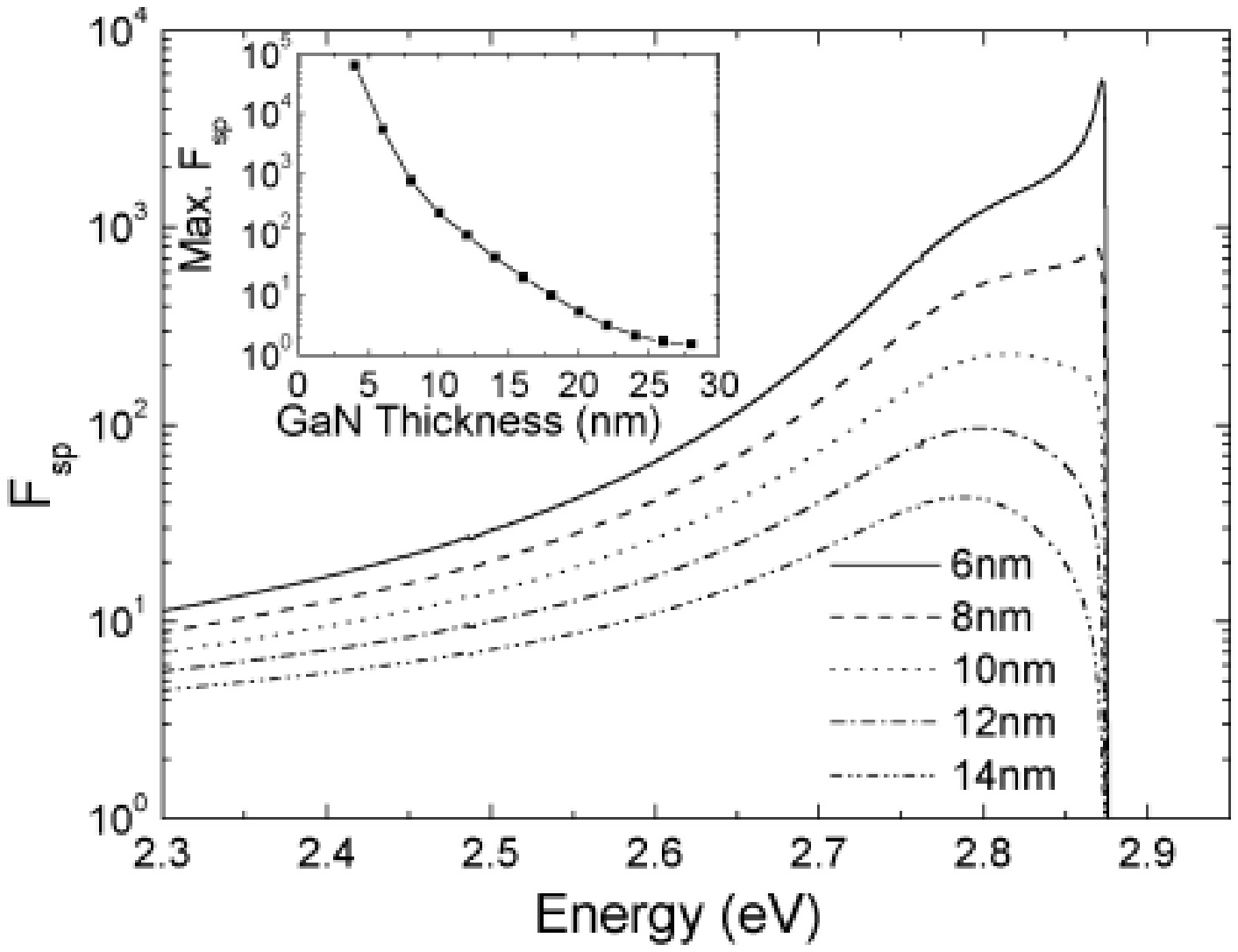}}} \caption{a)
Calculated $F_{sp}$ for various Ag thicknesses $t$ with Ag-QW
separation $a=$ 12 nm. Inset: Maximum $F_{sp}$ values for a given
$t$ with $a=$ 12 nm. b) Calculated $F_{sp}$ for various Ag-QW
separations $a$ with Ag thickness $t=$ 8nm.
 Inset: Maximum $F_{sp}$ values for a given $a$ with $t=$ 8nm.} \label{Fig4}
\end{figure}

Using (4) and (6) to estimate $F_{sp}$, even larger enhancements
are predicted over narrower frequency bands. For a prominent
resonance in $F_{sp}$, Fig. 4 indicates that it is necessary for
the Ag film to possess a thickness $\geq$ 6 nm for a QW 12 nm from
the surface. In support of the earlier deduction that the Ag film
is slightly thicker at $P_{2}$ than $P_{1}$, the value of $F_{sp}$
is predicted to increase with increasing film thickness and
asymptotically approaches 422 for $t\geq$ 26 nm ($a=$12 nm).

The strength and frequency dependence of $F_{sp}$ are even more
sensitive functions of the Ag-QW separation, especially for small
$a$.  Resonant enhancements of more than $10^{4}$ are predicted
for QWs only 4 nm below the surface.  These predicted enhancements
may actually be conservative because the synergistic "back action"
coupling between dipole emitters and SP field, which increases as
$a$ decreases, is not included in this calculation. A more
comprehensive calculation of this enhancement factor, including
the necessary radiation reaction effects, is beyond the scope of
this paper. Nevertheless, the enhancement will likely remain broad
because inhomogeneous broadening ($\hbar\Delta\omega_{inh}\simeq$
100 meV) will probably limit $\frac{\omega_{c}}{\Delta\omega}<$
30.
\begin{acknowledgments}
The authors would like to thank S. Kellar, U. Mishra, and S.
DenBaars of UCSB for providing the QW sample, and I. Gontijo,
M. Boroditsky, and G. Khitrova for useful discussions.
This work was supported by
the Army Research Office and the National Research Council.
\end{acknowledgments}

% \bibliography{basename of .bib file}
\end{document}